\documentclass[natbib, acmsmall, screen, authorversion]{acmart}

\usepackage[T1]{fontenc}
\usepackage[english]{babel}

\usepackage{mathtools}
\usepackage{mathrsfs,bm}
\usepackage[utf8]{inputenc}
\usepackage{xspace}
\usepackage[tikz]{mdframed}
\usepackage{makecell}
\usepackage{newfloat}
\usepackage{caption}

\usepackage{enumitem}
\setlist{noitemsep,leftmargin=!}

\usepackage[nameinlink,capitalize]{cleveref}
\addto\extrasenglish{%

}

\usepackage{boxedminipage}

\title{A Survey on Parallelism and Determinism}
\thanks{Partially funded by French ANR CODAS project (ANR-17-CE23-0004-01)}

\author{Laure Gonnord}
\orcid{0000-0002-8013-1611}
\affiliation{
  \institution{LCIS (UGA, Grenoble INP)}
  \postcode{26902}
  \city{VALENCE Cedex 09}
   \country{France}
}
\affiliation{
  \institution{LIP (EnsL, UCBL, CNRS, Inria)}
  \postcode{F-69342}
  \city{LYON Cedex 07}
  \country{France}}
\email{laure.gonnord@esisar.grenoble-inp.fr}

\author{Ludovic Henrio}
\orcid{0000-0001-7137-3523}
\affiliation{
  \institution{CNRS, EnsL, UCBL, Inria, LIP} 
  \postcode{F-69342}
  \city{LYON Cedex 07}
  \country{France}}
\email{ludovic.henrio@cnrs.fr}

\author{Lionel Morel}
\orcid{0000-0002-0246-1930}
\affiliation{
  \institution{Univ Lyon}
  \institution{INSA de Lyon, CITI Lab}
  \postcode{EA3720}
  \city{Villeurbanne}
  \country{France}}
\email{lionel.morel@insa-lyon.fr}

\author{Gabriel Radanne}
\orcid{0000-0002-2107-7678}
\affiliation{
  \institution{Inria, EnsL, UCBL, CNRS, LIP} 
  \postcode{F-69342}
  \city{LYON Cedex 07}
  \country{France}}
\email{gabriel.radanne@inria.fr}

\setcopyright{acmlicensed}
\acmJournal{CSUR}
\acmYear{2022} \acmVolume{55} \acmNumber{10} \acmArticle{210} \acmMonth{10} \acmPrice{15.00}\acmDOI{10.1145/3564529}

\newcommand\PLUS{{\color{Green}$\bm+$}}

\newcommand\MINUS{{\color{Red}$\bm-$}}
\newcommand\MEH{{\color{Orange}$\bm\pm$}}

\newenvironment{nicebox}[1]
{\begin{mdframed}[backgroundcolor=#1,linewidth=0pt,roundcorner=10pt]}
{\end{mdframed}}

\definecolor{strongpointsColor}{rgb}{0.8,0.95,1} 
\newenvironment{strongpoints}[1]%
{\vspace{0.6em}\begin{nicebox}{strongpointsColor}
    \begin{minipage}{\linewidth}
      {\bfseries\upshape Strong Points for #1:}
      \begin{itemize}
        [leftmargin=10pt,topsep=2pt, partopsep=0pt,itemsep=1pt,parsep=1pt]}
{\end{itemize}\end{minipage}\end{nicebox}}

\definecolor{DefinitionColor}{rgb}{1.0,0.9,0.92} 
\DeclareFloatingEnvironment[fileext=deff,placement={!tb},name=Definition]{deffloat}
\crefname{deffloat}{Definition}{Definition}
\newenvironment{definition}[2][tb]{
  \begin{deffloat}[#1]
    \begin{nicebox}{DefinitionColor}
      \caption{#2}}%
{\end{nicebox}\end{deffloat}}

\definecolor{ExampleColor}{rgb}{0.8,1,0.7}
{\begin{nicebox}{ExampleColor}}%
{\end{nicebox}}

\definecolor{carmine}{rgb}{0.59, 0.0, 0.09}

\definecolor{darkgreen}{rgb}{0.1, 0.5, 0.1}

\begin{document}

\begin{abstract}
  Parallelism is often required for performance. In these situations an excess of non-determinism is harmful as it means the program can
  have several different behaviours or even different results.
  Even in domains such as high-performance computing where parallelism is crucial for
  performance, the computed value should be deterministic.
  Unfortunately, non-determinism in programs also allows dynamic scheduling of tasks, reacting to the first task that succeeds, cancelling tasks that cannot lead to a result, etc.
  Non-determinism is thus both a desired asset or an undesired property depending on the situation.
  In practice, it is often necessary to limit non-determinism and to identify precisely the sources of non-determinism in order to control what parts of a program are deterministic or not. 

This survey takes the perspective of programming languages, and studies how programming models can ensure the determinism of parallel programs. This survey studies not only deterministic languages but also programming models that prevent one particularly demanding source of non-determinism: data races.

  Our objective is to compare existing solutions to the following questions: How programming languages can help programmers write programs that run in a parallel manner without visible non-determinism?
  What programming paradigms ensure this kind of properties?
  We study these questions and discuss the merits and limitations of different approaches. 
\end{abstract}



\maketitle




\newpage



\section{Introduction and Scope of the Survey}

\subsection{Parallelism: why and how?}
In computer science, parallelism relates to the simultaneous execution of different tasks contributing to the same application. Parallelism can either be expressed in programs' source code or be introduced during their compilation or their execution.

An application may be parallel by nature because its functionality needs to be performed by several distinct computing units, for example if it gathers information originating from different geographical locations. Other applications are perfectly described in a sequential manner but parallelism can be introduced in the program, its compilation, or its execution in order to increase its overall efficiency or responsiveness. Programs can be split into different tasks that can run in parallel; these tasks can either fulfil different roles, e.g. computation vs. input/output, or handle different parts of the work to be performed. 

Two very common abstractions proposed are \emph{task-parallelism} and \emph{data-parallelism} (see \cref{def:TaskDataParallelism})\footnote{The fundamental terminology for understanding the survey has been highlighted and defined in pink boxes. Most of the definitions are ``common knowledge'' in the programming language community but some alternative definitions exist, this is why  it is important to clearly state the definitions we use.}.
Task-parallelism consists in splitting the application's work into several  tasks to be performed. These tasks can  execute in parallel and  well-specified events can be used to coordinate their execution. 
This programming paradigm is particularly adapted when the tasks are different and relatively complex.
Data-parallelism comes from an identical treatment (e.g. the exact same function is used) being applied to different pieces of the input data. In that case, communication and synchronisation are used to ensure coherence of treatment over the different parts of the data treated in parallel. 
While the two models are expressive enough to express all kinds of parallel programs, programming a task-parallel problem as a data-parallel pattern or the contrary is cumbersome and inefficient.

\begin{definition}[b]{Task parallelism and data parallelism}
  \label{def:TaskDataParallelism}

\textbf{Task parallelism}: A parallelisation paradigm that consists in splitting the application into several tasks to be executed. Parallelism comes from the simultaneous execution of tasks.

\textbf{Data parallelism}:  A parallelisation paradigm that consists in providing a single task to be performed on a set of (homogeneous) data. Parallelism comes from the fact that the data is split and different instances of the task treat the different data parts.
\end{definition}

\subsection{Parallelism and Determinism}

%

When tasks run in parallel, their effect may lead to a conflict, such as two threads trying to write data at the same time to the same memory location, or two threads trying to be scheduled concurrently.
Such conflicting behaviours create non-determinism that can lead to  different executions with different behaviours, and thus make the programming more error-prone and difficult to debug.
Additionally,  determinism properties can improve both  execution support and debugging. For example, partial determinism properties of active objects have been successfully used to design recovery protocols in the context of fault-tolerance~\cite{BCDH:PPOPP2007}, and to design deterministic replay techniques in the context of debugging support~\cite{TJS:FASE2020}. 

Let us consider the different behaviours of a program.
According to  the rewriting theory or the foundations of programming language semantics, a program (or a semantics) is  deterministic if there is a single way to execute it and confluent if it can be run in different manners but all executions (scheduled differently) globally have the same result.
Monitoring a confluent program leads to different traces, which is not the case for a  deterministic program.
If the execution is also finite confluence means that all executions
lead to the same final state.
If the execution is not finite, confluence means that, from any two intermediate states reached by two partial executions, there is a way to schedule the rest of the executions  so that they both reach a common (intermediate) state~\cite{baadernipkow:book98}.

 Because most of the works we study here use the words determinism and confluence interchangeably, in the following we use the term ``\textbf{determinism}'' for both concepts of rewriting theory: determinism and confluence.
This terminology comes from the observation that both deterministic and confluent rewriting systems produce a deterministic result. Thus the difference between the two concepts is not useful for the programmer.

Non-deterministic programs exhibit \emph{race-conditions}, see \cref{def:Races}. 
To tame the difficulty to write and analyse non-deterministic programs several concurrent programming languages limit the possibility to have races. 
The absence of data races is an interesting  property because it ensures that non-determinism only originates from conflicting communications or synchronisations.
Data-race freedom and determinism properties are often ensured at a language level, e.g. all programs using actors are data race free  or all dataflow synchronous programs are deterministic.

We call \emph{partial determinism} any property that identifies the conditions under which the execution of a program leads to a deterministic result.
For example, based on the absence of data-races, some programming models are able to ensure partial determinism. 
One instance of such a partial determinism property has been shown in the context of ASP (Asynchronous Sequential Processes)~\cite{CHS:POPL04}: in an actor model with a FIFO service policy and futures with blocking reads, non-determinism only originates from two conflicting communications toward the same mailbox.
More generally, several static analyses, compilation approaches, or runtime frameworks also ensure partial determinism properties.

In this paper we classify the approaches based on the ``programming model'' they rely on. We use programming model to designate the concepts manipulated by the programmer and their semantics. 
Because the programmer's main entry point to development is the
programming model itself, we have chosen to focus our presentation on
this criterion.  Rather than a \textit{language} vision, we consider
\textit{programming models} as a distinguishing factor in order to
highlight the fact that important criteria for the choice of a language should not relate
 on syntactic details of a real programming language but more
on the programming concepts that are exposed.

This paper is written with two types of readers in mind: \textit{application programmers} that desire to pick the
right approach with respect to their needs; and \textit{language designers},
that wish to provide better safety guarantees or more parallelism in their languages. As
opposed to previous surveys~\cite{parallelsurvey98,Kessler2007ModelsFP} that
mainly focus on cost models and performance, we focus on programming approaches that provide partial determinism properties. In other words, the focus of this survey is more on safety than on performance.

\begin{definition}{Races}
  \label{def:Races}

\textbf{Race-conditions}: A \emph{race-condition} occurs when two actions are possible and conflict in the sense that they lead to different global executions.
 
\textbf{data races}: A \emph{data race} is a particular case of race-condition where the conflicting actions are the writing or reading of data at the same memory location.
\end{definition}

The rest of the paper is organised as follows. In~\cref{sec:welovelanguages}, we describe the scope of our study: we focus on programming models and their safety properties regarding parallelism.
The next two sections explore various parallel programming models of the literature, classified in two families: the deterministic ones in~\cref{sec:confluence} and the data race free ones in~\cref{sec:dataracefree}.
Finally, we summarise our study in~\cref{sec:summary} where we compare the different programming models, exhibiting how they can be decomposed in complementary building blocks, each of which serves more specific purposes leading to generally-speaking deterministic programs.
0

\section{Positioning: Language and Program Analysis for Determinism}
\label{sec:welovelanguages}

Deterministic parallelism can be addressed at different steps of the software life-cycle. These different steps are briefly described below, we also explain how each step can be used to produce efficient and correct parallel applications. 

\subsection{The Different Phases of the Software Life-cycle}

\paragraph{Programming} Many programming languages or libraries let the programmer declare the intrinsic or potential parallelism of his/her application.
Some primitives in the programming language are sometimes intended for the programmer to express this potentiality. It is then the responsibility of the programmer to state which parts of the program can be parallelised safely and efficiently. 

\paragraph{Typing and Static Analysis}
Type systems and static analyses serve two purposes. First, they reject programs for which there is no well-defined behaviour. Second they are used to extract properties about programs. These analyses can be used to check that the parallelism declared by the programmer is correct, or to extract potential parallelism, e.g. by inferring that two parts of the program can be separated into independent entities running in parallel.

\paragraph{Compilation}

Modern hardware has so many characteristics that fine-grain optimisations are often intractable for the programmer.
Compilers are often used to explore possible transformations and generate code that a human cannot imagine by herself.
In particular, they deal with  programming issues which should remain hidden to the programmer.
In the field of domain specific languages, even more details need to be kept hidden from the programmer, either because she is not a developer or because we want her to focus on the modeling of specific artefacts such as the external environment (in reactive systems) or temporal properties (in critical systems). 

Compilers are complex pieces of software that rely on a large set of static techniques to generate code.
The efficiency of these analyses and transformation techniques is linked to the programming language and the static analyses, especially when it comes to parallelism extraction. For example, a language with strong parallel constructions needs less analysis effort to be compiled into optimised parallel code.

\paragraph{Runtime}

Runtimes play a role  whenever a decision such as code placement, code scheduling, or data distribution cannot be made statically or can only be performed with substantial over-approximations.
For instance, these decisions may highly depend on dynamic data, or on hardware components  shared with other programs running on the system.
In this case, the runtime (be it the operating system or a language-specific interpreter or runtime, e.g. openMP's runtime) can implement such decisions. 

\paragraph{Hardware}

Modern hardware often itself performs clever optimisations like branch prediction. Every such optimisation is by essence more efficient than the corresponding software version. However, the complexity of modern architectures makes predicting the exact impact of these optimisations difficult, and making a clever usage of hardware characteristics is still a challenge for all the steps above. 

Also, many hardware designs were proposed that could be naturally mapped to some parallelism-oriented programming models. In particular, dataflow processors~\cite{dataflowprocessors99} are still  extensively explored as a way to efficiently execute applications following this programming models.
We believe these hardware designs represent a relatively small niche and we rather concentrate on programming models that are executed on the more classical Von Neumann-style processors.


\subsection{Language, Compilation, and Program Analysis Approaches}
Programming languages are not only designed to allow the programmer to express computations or programs efficiently. They also allow safety verification of programs: they let the programmer provide the right information to a static analysis and to the compiler to enable efficient and safe execution.
In particular, high level programming abstractions give a high level point of view to the programmer; the important global information on programs is directly expressed by the programmer and does not have to be inferred through static analysis of lower-level constructs.

There are different approaches to express parallelisation in a program: classical full-fledged languages, domain specific languages, parallelisation libraries, pragmas, and annotations. These approaches have similar characteristics but with different guarantees, tools, and development costs.

\textbf{A full-fledged language} is a very flexible approach where the designer of the language can introduce rich constructs with powerful constraints on the program,  ensuring by construction strong guarantees on the executed application. The drawback is that developing and maintaining a language costs a lot of time and energy. A frequently used solution to mitigate this cost is to compile to a low-level language, typically C~\cite{ref:encore15} or sometimes to richer languages (e.g., ABS features different backends including Haskell, Maude, or Erlang~\cite{JHSSS10}), for which an efficient compiler already exists.

\textbf{Domain specific languages (DSL)} are languages that target a specific domain and generally compromise expressiveness for dedication through specific operators or programming structures. The DSL approach generally relies on a similar approach as full-fledged languages but allows the language developers to spend most of their energy in the efficient compilation of constructs that are specific to the targeted domain. 

The previous approaches rely on the implementation of a compiler
(implementing the semantics defined by the language, generating machine
code that fits this semantics); in the case of parallelism, compilers
can also reschedule, distribute, or merge parallel code.  
To avoid having to develop and maintain a specific compiler, one alternative is to rely on an existing language and extend it.

\textbf{Developing libraries} is the least intrusive way to do this. What can be expressed as a library is easier to maintain and a library may provide enough constructs to allow the programmer to express rich parallel programs. Writing complex libraries is easier when the host language provides enough information on the programs themselves. Indeed complex libraries often rely on meta-programming and reflection~\cite{AkkaBook,CDD:CMST06} (i.e. the ability to manipulate some information about the program itself). The main drawback of the library approach is that it often relies on ``good practices'' because the library itself is not able to enforce strong properties on the program, especially if the programmer uses features of the original language that conflict with the purpose of the library. For example, \cite{Tasharofi2013} analysed several Scala projects where programmers mixed the actor model with other types of concurrency in a generally unsafe manner, threatening the absence of data-races and the (partial) determinism properties of actors.

Finally, \textbf{pragmas and annotations} allow the programmer to attach information to an existing program. Some of this information can be used to parallelise the program, sometimes changing the program behaviour because of the concurrency introduced. Annotations can also be viewed as a way to control the amount of concurrency introduced. Annotations are the core principle of the standard openMP approach~\cite{Dagum1998}. They have also been introduced, and particularly well integrated with the concepts of the language, in Jac, a Java framework for concurrency~\cite{lohr2006jac}.  Pragmas differ from libraries in terms of expressiveness: the language expressed inside the annotations can be adapted to the target domain compared to the general-purpose host language, but the connection between elements of the host language and elements of the annotations might be complicated. Additionally, what can be expressed inside the annotation language needs to be designed carefully to reach the right level of expressiveness.

Complementarily to these different views on programming model approaches, static analyses can be designed in order to
infer some properties of parallel programs such as absence of
deadlock or race conditions~\cite{racerx,mine:hal-01105235,GHLM-PPDP16}. These analyses can be
performed on any type of language, from general purpose languages with
explicit threads and mutexes, to languages that propose high-level constructions like futures. 

\smallskip

All these approaches might be combined in many ways  to provide the desired language features. In this survey, we focus on programming paradigms that provide at least data race freedom, the next section details what is inside the scope of this survey and what are the related works outside the strict scope of this survey.

\subsection{Scope of the survey}

In this survey, we take a programming language approach on the design of deterministic or data race free programs. 
Before getting into the details of partial determinism in the context of programming models, we summarise in this section alternatives approaches and surveys on related topics. This section thus focuses mostly on approaches that either do not ensure the absence of data races, or do not rely on the definition of a  programming model.


The choice to take a programming model perspective on data race freedom is particularly relevant because, on the programming model side, there is a clear distinction between
data race free and non-data race free languages.
The focus of our survey is on programming models where data race freedom is part of the language philosophy.
This means that the ``obvious'' code using these languages should be data race free, even if there might exist some out-of-the-way escape hatches or workaround to create data races.
Similarly, we include in our survey a section on linear types (\cref{sec:lineartypes}) because it is a language construct that by nature ensures data race freedom, even if there are languages with linear types that are not data race free.
On the contrary, we do not cover languages where data race freedom is only achieved through alternative tools such as static analyses.
We briefly describe these out-of-scope programming paradigms in the next paragraph.

\paragraph{A few classical parallel programming models}
Historically, numerous parallel libraries have been designed to enhance the parallel and distributed capacities of C programs in addition to classic threads. 
First, the classical OpenMP library~\cite{dagum1998openmp} is composed of liberal constructs that do not prevent data races, even if \texttt{critical} and \texttt{atomic} directives might help to design safe programs.  Nevertheless some attempts do exist to detect data races in OpenMP programs~\cite{openmp:races2012}. 
The Cilk language~\cite{cilk2009}, a multithreaded parallel C-like language, similarly has data races which can be avoided using specially designed data structures called \texttt{hyperobjects}.
X10~\cite{X10} is a similar Java-based language where static analysis is necessary to detect tasks that can occur in parallel and cause data races~\cite{ABSS07:X10races}.
The MPI library~\cite{MPI} aims at providing communication and distribution primitives for programs to be executed on large clusters. Its message-passing based approach does indeed helps the programmer to design data race free programs. We however exclude it since, compared to the languages we study here, data race freedom relies more on the coding practises than on the language itself. For instance, MPI communications can be made efficient by using Remote Memory Access, which mandates the use of static analysis to ensure the absence of data races~\cite{MPI-RMA-dataraces-2021}.

\paragraph{Transactional Memory}

Transactional Memory is a \emph{runtime} approach to provide optimistic
lock-free constructs to access and modify memory simultaneously.
Each agent wanting to access the
shared memory locations will attempt to execute and commit a transaction.
One will succeed and apply
its operations, while the others will abort and may retry.
Transactions were originally intended to use dedicated hardware support
\cite{DBLP:conf/isca/HerlihyM93}, but were soon implemented
in software \cite{DBLP:conf/podc/ShavitT95} and now support a wide
range of techniques and properties.

Transactional memory approaches provide limited control over interleaving,
by ensuring
transactions behave similarly to critical sections, while still allowing
some runtime concurrency for efficiency. This relies on some relaxation of
determinism properties.
For instance, these approaches can provide various properties~\cite{DBLP:journals/toplas/HerlihyW90}
such as ``obstruction-freedom'' (absence of deadlock but allow livelocks)
or ``lock-freedom'' (absence of both deadlocks and livelocks).
Concretely, they provide neither full determinism
nor data race freedom for the whole language, but rather rely
on \emph{opacity}~\cite{stm:correctness2008} which means that
from an external view point, different transactions should behave
as if they had been executed sequentially and always have a consistent view of the memory. As a consequence, if the programmer encapsulates every critical section inside transactions, no data-races are possible.

\paragraph{Related surveys}


A first related survey can be found in \cite{replay:survey2015} concerning
deterministic replay, which is very useful for debugging or
fault-tolerance. Replay
techniques enforce one execution of a program to be identical to a previously recorded execution. These techniques thus mostly apply to programming languages
that are not by nature deterministic, potentially with data races and other sources of non-determinism. 
On the contrary, we focus here on approaches that allow the programmer to  write (almost) deterministic programs directly, thus easing their understanding and debugging. Note that partial determinism properties of programming models can be used to design efficient deterministic replay techniques as less information needs to be logged to characterise an execution~\cite{CH-book}.
In addition, a significant part of the survey \cite{replay:survey2015} is
placed on lower-level aspects (operating systems, hardware, etc.)
which are significantly different from the language point of view we
adopt. For instance it studies deterministic replay in presence
of weak memory models or data races, which is clearly out of our scope
here.

Similarly, thread-based speculation
techniques~\cite{speculation:survey2017}, allow running originally sequential piece
of code in parallel. Dependence violations are detected at
runtime (either with hardware of software mechanisms) and roll-back
mechanisms are used to replay the sequential code if required. These
techniques can thus be viewed as a way to increase parallelism for deterministic
programs without introducing non-determinism.  
Even if they can be complemented
by language mechanisms that enables dependencies to be efficiently
computed, we consider these approaches out of scope.

Finally, other surveys about programming parallel machines have been published in the past.
The present article can be viewed as a follow-up to the
survey~\cite{Kessler2007ModelsFP} about models of computations: our
``programming model'' focus benefits from more recent contributions
that have transformed the ``general parallel programming
methodologies'' described in the 2007's paper into what we call ``programming
models/languages features''.
More recently, \cite{Diaz2012} present a rather thorough survey
on programming languages covering pure parallel models (OpenMP and
MPI), heterogeneous models (CUDA, OpenCL), Partitioned Global Address Space models,
as well as models mixing several of these approaches. Compared to \cite{Diaz2012}, our
survey has a narrower focus, especially on guarantees brought by
programming models about determinism.

\subsection{How Programming Models Limit Non-determinism}
\label{sec:limitnondet}

From the study of several deterministic and partially deterministic languages, we identified the main solutions that exist to limit the non-determinism of parallel programs. The main part of this survey will present the  languages that we considered as particularly interesting in this context.
As stated above, in many cases, non-determinism is an undesired
property, and consequently several languages provide abstractions
restricting or forbidding non-determinism behaviours.
In the sequel, we call ``action'' every single possible transition of
our programs (a ``step'' in the terminology of operational semantics, or an ``event'' in the distributed system terminology). Actions can be for example a data read or write, a communication through a message, a thread creation. Two
actions are \emph{racy} if they are both enabled at the same instant
of execution, and the observable remaining behaviour of the system
is different depending on which of the two actions is executed
first. The fact that we are interested in the observable differences  introduces  a notion of equivalence where we only consider the visible result of the execution. 
A language is called
\emph{deterministic} if there cannot be any racy action. A racy action can be formally identified by studying, for example, the operational semantics of the language.

As we are interested in  solutions for
parallelising an application while  producing a
deterministic result, these solutions should thus avoid  racy actions. 
These races between actions can be characterised as:
\begin{description}

\item[Data races] Two actions  that read or write to the same memory location can be racy if one of the two is a write. Programming safely with data races
  is difficult and data races are avoided in most of the partially
  deterministic programming models. Additionally, in the presence of weak
  memory models \cite{DBLP:journals/computer/AdveG96}, the impact of data races is often more difficult to
  predict. All the programming models presented in this survey will
   at least strongly limit data races thanks to additional
  constraints such as single assignment (\cref{def:sassign}), blocking read (\cref{def:blockingread}), or partition of the memory. 
Because data access is well-protected, the programming models described in this survey are not sensitive to the reordering of memory accesses. This is why weak memory models~\cite{DBLP:journals/computer/AdveG96} are almost absent in the state-of-the-art of the models presented here.

\item[Synchronisation and communication races] When two actions that synchronise two threads can be executed at the same time and are exclusive, like a lock release or the execution of a critical section by two entities, a race occurs. 

  Similarly, if the same channel can be used by two different message
  sending actions, or two different message reception actions then
  these two actions constitute a racy condition. Communication races can be arbitrarily complex depending on the complexity of the communication
  patterns. Communication races can be limited by providing
  constraints on the type of communication or their schedule, e.g. by
  imposing FIFO communication over channels (\cref{def:FIFO}).

\end{description}

The  categories cited above are not  disjoint, in particular  some actions can be considered both as memory access and communication, or both as synchronisation and memory access.  This is the case for example to shared memory accesses in distributed systems; whose concurrent reads may be considered as data or communication races.

\medskip

The languages we study in this survey limit the potential races in different ways. These solutions can be classified as follows:
\begin{itemize}
\item \textit{Limiting the semantics} so that no conflicting actions can be observed as in synchronous languages or in purely functional languages;
\item \textit{Limiting the interactions between computing entities} to prevent races as in actors, or BSP (Bulk Synchronous Parallel) \cite{Valiant1990} which prevents races between computation and communication (in two isolated phases);
\item \textit{Providing communication and synchronisation tools} that are \textit{by nature data race-free} like futures or Single Assignment;

\item \textit{Restricting communication scheduling} to limit the set of potential behaviours and prevent some communication races (FIFO channels / mailboxes, see~\cref{def:FIFO});
\end{itemize}

\begin{definition}{Single Assignment}
  \label{def:sassign}
  One way to get rid of write/write conflicts is to forbid multiple writes to the same (memory) location: such a pattern is called ``Single Assignment''~\cite{ssa88,ssa91}.
\end{definition}

\begin{definition}{Blocking Read}
  \label{def:blockingread}
  A read to a location is said to be \emph{blocking} if the reading process is suspended until this part of memory actually contains (updated) data.
\end{definition}

\section{Programming models with determinism guarantees by construction}
\label{sec:confluence}

Several programming models can somehow reach the goal of combining parallelism and determinism (with some restrictions on the form of parallelism), they are presented in this section. Many of the programming models presented in this section are based on synchronous  or dataflow programming models.


\subsection{Kahn Process Networks}
\label{sec:kpn}

The first dataflow programming models (see~\cref{def:dataflow}) were proposed in the seminal work of G.  Kahn~\cite{Kahn1974}. Those constructs were later called KPNs, for Kahn Process Networks. 
At the heart of KPNs is the idea to decompose the application as a set of independent processes or \textit{agents} that communicate solely through First-In-First-Out (FIFO) channels.

\begin{definition}{FIFO channel}
  \label{def:FIFO}
  A \textit{FIFO channel} (also refered to as \textit{FIFO queue}) is a data buffer with exactly \textit{one} producer  and \textit{one} consumer, where data tokens are forwarded to the consumer in the exact same order they have been emitted by the producer.
  A FIFO channel can hold a varying number of elements, and these elements are given exactly \textit{once} to the consumer.
  Reading a FIFO is \textit{blocking} (see~\cref{def:blockingread}): when the consumer  tries to read an empty FIFO, it is suspended until at least one data element is emitted by the producer.
  FIFO channels are usually \textit{single assignment} (see~\cref{def:sassign})  while writing (they forbid multiple rights to the same cell). 
  Depending on the programming model, FIFO buffers can be considered to hold an unlimited number of data tokens.
 Emitting on  a FIFO channel is most often a non-blocking action. 

Note that an \textit{actor} (see \cref{sec:actors})
 may be equipped with FIFO \emph{mailbox}es. Such a mailbox is not a channel \textit{per se} because it supports several producers.
\end{definition}

%
%
%
%
When a KPN agent is scheduled for execution, there is essentially no way to know how many data elements it will read/write from/to its input/output channels. 
As a consequence, buffer sizes as well as scheduling of agent's activations cannot be determined statically.
Also, no guarantee can be given that a KPN program can be executed with finite memory.
It is therefore the task of a dynamic scheduler to attend to these questions.
As an example, consider a video encoder that processes constant-size blocks of raw images and produces compressed encodings corresponding to these blocks.
Statically, there is no way to know the size of blocks produced by the encoder as it essentially depends on the content of each block in the input raw image.
While the global correctness is easy to establish, the size of the communication channels cannot be predicted statically.
In some contexts, like critical control systems, this is unacceptable, as static guarantees on the necessary memory are needed to ensure the system's safety.
However, in many contexts, KPNs offer the right level of expressiveness because of the simple and deterministic parallelism they provide.
%
%

%
To address the question of static predictability of boundedness, many variant models and languages have been studied. They focus on the ability to determine at compile-time how many data tokens are exchanged by agents over FIFO channels.
These \textit{static dataflow programming} models are quickly reviewed now.

\begin{strongpoints}{KPN}
  \item[\PLUS] Very fertile ideas that have been used as inspiration for many parallel languages.
  \item[\PLUS] Simple semantics, and simple syntactic requirements to ensure determinism. 
  \item[\MINUS] Restricted programming patterns and parallelism. 
  \item[\MINUS] Unpredictable size of channels. 
\end{strongpoints}

\begin{definition}{Dataflow}
  \label{def:dataflow}

  A \textit{dataflow} programming model~\cite{karpmiller:df66,Kahn1974} is a model where a program is decomposed into actors that solely communicate through FIFO channels. 
\end{definition}

\subsection{Static Dataflow Programming}
\label{sec:sdf}

In Static DataFlow Programming (SDF in short)~\cite{Lee1987:sdf}, agents communicate through channels. The programmer gives, for each agent, the (static) number of tokens it reads (resp. writes) on each of its input (resp. output) channels, each time it is activated.
%

Static dataflow programming models focus on properties necessary to ensure boundedness of memory usage as well as static schedulability. In fact, the problem of statically scheduling agents is shown to be decidable, for sequential as well as for parallel (homogeneous) processors (with maximum parallelism). 
%

The $\Sigma$C language~\cite{cea:emsoft12,streamsigc}, which implements the Cyclo-Static Dataflow model~\cite{Bilsen1996} where data rates
of agents are known statically and can change periodically, has also the same property.  Depending
on the target architecture, FIFO communication channels are compiled
into efficient shared-memory implementations or into distributed
communication mechanisms. 
The $\Sigma$C runtime  is then in charge of allocating  memory regions and
scheduling the activities corresponding to  agents, either relying
on an underlying operating system or through dedicated scheduling policies.

Parallelisation of Static Dataflow programs has been well studied for the StreamIT language~\cite{Thies2009,Gordon2010}.
The StreamIT compiler and runtime manages both data- and task- as well as \textit{pipeline}-parallelism. 
Also, StreamIT distinguishes explicitly between stateless agents, which can be trivially duplicated for data-parallelisation and stateful actors, that embed a state, which makes parallelisation possible in certain cases but much trickier~\cite{Schneider2012}. 

%
%

\begin{strongpoints}{SDF}
\item[\PLUS] Determinism inherited from the KPN model.
\item[\PLUS] Static scheduling and statically bounded channels.
\item[\MINUS] Requires static information on communication rates given by the programmer.
\item[\MINUS] Restricted parallelism patterns (similar to KPN).
\end{strongpoints}

\subsection{Synchronous Languages}
\label{subsec:sync}

\newcommand{\lustre}{{\sc Lustre}\xspace}
\newcommand{\esterel}{{\sc Esterel}\xspace}
\newcommand{\signal}{{\sc Signal}\xspace}

\begin{definition}{Synchronous Language}
  \label{def:sync}
  A \textit{synchronous language}~\cite{Halbwachs1991,leguernic:signal03,esterel:berry} is a language dedicated to programming \textit{reactive systems}.

  It enforces a logical vision of time where on each tick of a \textit{clock}, the program receives new values for all or part of its inputs, processes these values and generates new values for all or part of its outputs. These output values are made available before the next tick of the clock. 
\end{definition}

Synchronous languages (See~\cref{def:sync}) like \lustre~\cite{lustre:halbwachs} have been proposed in the early 80s for the reliable development of safety critical embedded systems.
In \lustre, a program is described as a composition of components that communicate through data flows.
Each component is a function over streams of data tokens.
Computations and communications are decoupled from one another: whenever a component is ``activated'' (i.e. scheduled), all its inputs are available. It then produces all its outputs immediately to its output environment. 

From the global system's point of view, the relations between components through their communication channels yield a partial occurrence ordering of all observed events: while focusing on a node, all its local events (used to compute its outputs) are totally ordered with respect to its abstract clock.
The language is deterministic, and the sequential compilation process of synchronous language~\cite{Halbwachs1991} generates code that is a correct sequentialisation of every computation of every node.

\paragraph{Parallel execution of Lustre programs}
Overall, while Lustre expresses deterministic execution as a set of independent
entities,
the efficient parallel execution of \lustre programs is still a challenging task.
Indeed, as stated in the survey~\cite{dist-sync-survey}, the problem is not to exhibit the greatest possible parallelism in the synchronous source program, but rather to deploy it onto a given (distributed/parallel) architecture, with some fixed level of parallelism. Approaches that address this problem include replication of the control and/or partitioning algorithms to distribute tasks~\cite{interf-europar,distributed-esterel00,distributed-gals02}.
%
%
All these algorithms and approaches, even the most sophisticated ones, still guarantee the preservation of dependencies and the determinism of the overall application.
%
%
In the context of hard-real time critical systems, the work presented in \cite{DBLP:journals/tecs/CaspiSST08} proposes a completely safe multitask implementation of \lustre parallel programs that is compatible with well known real-time scheduling algorithms like fixed-priority or earliest-deadline first: 
the authors of \cite{DBLP:journals/tecs/CaspiSST08} prove the preservation of synchronous semantics by implementing communication channels as buffers with concurrent accesses.


\begin{definition}{Future}\label{def:fut}
A \emph{future}~\citep{BakerHH77,Halstead85,jlambda-fut06} is an entity representing the 
result of an ongoing computation.  It is  used to launch a sub-task in parallel 
with the current task and later retrieve the result computed by the sub-task. 

By nature, futures are single-assignment entities that can support multiple readers. 
Several semantics for synchronising on the future resolution exist but if only blocking read is used then futures do not introduce non-determinism~\cite{CHS:POPL04}.
\end{definition}

\paragraph{Introducing asynchrony into synchronous languages }
Alternatively to the previous approaches whose common point is to preserve synchronicity as much as possible, the authors of~\cite{lustrefutures:emsoft12} suggest to extend \lustre with futures (see \cref{def:fut}), providing more asynchronism without losing determinism. 
This allows easier expression of non regular synchronisation, or fork/join programming models that are tedious in standard \lustre, albeit essential idioms for elegantly writing certain programs.
This is performed by the combination of a keyword \textit{asynch} that triggers asynchronous computations, and explicit blocking synchronisation on the future that corresponds to such asynchronous function executions.
The communication between the task that launches the \textit{asynch} and the \textit{asynch}ronously called one is done by a bounded FIFO channel with a statically predefined size.
%
%
Each of these entities can also be multi-threaded to have several processes perform the same task in parallel, but  only  if the task is stateless (see \cref{sec:sdf}).
The approach is implemented as a fork of Heptagon~\cite{delaval:hal-00863286}, a language very similar to \lustre.
It is interesting to notice that the notion of future is also massively used in actors (see \cref{sec:actors}) and that the runtime used to execute \lustre programs with futures is similar to an actor system.

Some other variants of \lustre have proposed extensions or restrictions of clocks (e.g., the integer clocks of~\cite{gerard:hal-00728527}) that enable efficient and proven determinism at runtime.
Prelude~\cite{cordovilla:inria-00618587}, one of the most mature of these variants, has a compilation and scheduling process that guarantees the determinism of computations that are eventually executed in their runtime.
These solutions mostly emerged in the context of HPC where the strict synchronous vision is too restrictive, as many applications need relaxed forms of synchronisation. 

\begin{strongpoints}{Synchronous Languages}
\item[\PLUS] The programming model is a good fit for certain applications, notably embedded reactive systems.
\item[\PLUS] Static inference of scheduling and communications.
\item[\MEH] Tailored for task parallelism.
\item[\MINUS] Parallel efficiency is difficult to obtain.
\end{strongpoints}

\subsection{Polyhedral Model}\label{sec:seq}


The principle of the polyhedral model is to rely on dependence analysis in order to parallelise the execution of a (sequential) program that is automatically scheduled in a correct manner (from a data dependency point of view). The programmer thus writes a sequential program.

Because it features no parallelism, sequential programming is the obvious deterministic programming model, and the polyhedral ``model'' (framework) is one of the most elegant approach to execute a sequential program in parallel. The ideas of the framework date back  in the late 70s, when the need for more performance in physical simulations (weather forecast, physics simulations for various applications) has crucially pushed for efforts to take advantage of hardware parallelism. The contribution on computing dependencies, by Karp, Miller and Winograd~\cite{karp67}, the efforts on systolic arrays~\cite{systolic:journal84}, as well as the work on loop transformations initiated by Lamport~\cite{lamport74} made the foundations necessary to the emergence of the whole framework in the early 90s.

%
With the polyhedral model, iterations of compute-intensive loop kernels are abstracted away as integral points satisfying affine constraints, namely, \emph{polyhedra}.
This framework~\cite{Feautrier92part1} provides exact dependence analysis information where statement instances (i.e., statements executed at different loop iterations) and array elements are   distinguished.
The exact dependence information  and the use of linear programming techniques to explore the space of legal schedules~\cite{Feautrier92part2} is what  constitutes the basis of the polyhedral model for loop transformations.

The traditional use of polyhedral techniques in optimising compilers
typically focuses on loop transformations of \emph{polyhedral
  kernels}. PLuTo~\cite{bondhugula2008practical} is now widely used as 
a push-button tool for automatically parallelising polyhedral loop nests.
It tries to optimise locality in addition to
parallelisation.
Polyhedral techniques for loop transformations are
now adopted by many production level compilers, such as GCC, IBM XL,
and LLVM.  There is also significant work in data layout optimisation
for polyhedral programs where analyses are performed to minimise the
program's memory requirement~\cite{darte2005lattice}. Recently, more modular
approaches~\cite{derien:compaan08,streamppn,dpn:cc21} have
emerged  that promote the
generation of intermediate dataflow models. With these approaches,
the extracted parallelism is represented as \textit{process networks}
that communicate through FIFOs. This enables further optimisations to
be performed in a modular manner.

As for properties, all polyhedral techniques have the same
\textit{guarantees by construction}: the programs obtained after transformation
compute the same result (i.e. have the same semantics) than the one from the input sequential program.
Only valid programs are
generated, and the dependences of computation are preserved.
However,
to be effective, this theoretical result strongly relies on safe
implementation of communication
patterns~\cite{le:hal-00862450,dpn:cc21}.
%
Finally, recent works such as~\cite{basupalli2011ompverify, yuki2013array} propose a safety analysis of transformed parallel programs that guarantees the absence of races.
Beyond the \emph{all-static} approach of parallelising compilers, with a particular focus on new applications such as neural networks~\cite{tiramisu:cgo2019}, some hybrid approaches have also been proposed to enhance the expressive power of the polyhedral model. Among these recent works we can cite the sparse polyhedral model~\cite{sparsepoly:pldi2019} and loop speculation~\cite{lazcano:hal-02457425}.

\begin{strongpoints}{Polyhedral Model}
\item[\PLUS] Programming model is a good fit for HPC kernels.
\item[\PLUS] Efficient execution obtained quasi-automatically.
\item[\MEH] Tailored for data parallelism.
\item[\MEH] The programmer has many parameters to play with to control optimisations.
\item[\MINUS] Only applicable to well-formed computation kernels.
\end{strongpoints}

\medskip

In this section we have seen several programming models that  achieve determinism while allowing for parallel execution. To achieve this, the only solution  is to prevent concurrency between conflicting effects. We have seen several techniques that allow parallelism without non-determinism: specific communication patterns like FIFO channels with blocking read on channels; specific semantics that deals with time in particular ways, like synchronous semantics; compilation techniques that generate parallelism for sequential programs, like the polyhedral model.

\section{data race Free Programming Models}
\label{sec:dataracefree}
The programming models of the preceding section had strong properties but had a restricted applicability in the sense that they require a specific semantics that is sometimes far from the one of the mainstream programming languages. 
When aiming at a broader field of application, one solution is to rely on languages that limits the sources of non-determinism.
The most adopted solution that allows non-determinism in a controlled way is to only allow races concerning communications or synchronisations but to prevent data races.
In general, data races can lead to complex bugs and it is a good programming practice to avoid them. 
We review here the programming models that natively prevent them. 
In these languages, first the programmer does not have to worry about potential data races; and second some partial determinism properties can often be identified because non-determinism can only originate from some well-identified races between synchronisations or communications.

\subsection{Dataflow Programming Models that are not Fully Deterministic: DPN and CAL}
\label{sec:dpn}

Dataflow Processes Networks (DPNs) were introduced by Dennis~\cite{Dennis1974}. 
The CAL Actor Language~\cite{calspec2003} is an implementation of the DPN model.
It proposes a structure of actors that perform a sequence of steps, consuming tokens from input ports, computing, and then producing tokens onto output ports. Actors can be parameterised, which simplifies the design of programs while avoiding to duplicate code.
A set of CAL actors is compiled to a set of sequential functions which are scheduled dynamically by a dedicated runtime. 

CAL is not a deterministic language and according to the authors of the Caltrop specification~\cite{Eker01anintroduction}, an actor is specified by  \textit{firing rules} that possibly involve non-deterministic choice among the possible reactions. 
Some of these ambiguities are solved by the compiler while others are left to the scheduler's choice. 
For example, a programmer can build non-deterministic CAL actors by using overlapping conditions in case/switch statements.
However, the language was designed to enable tools to identify possible sources of non-determinism and warn the programmer about them.
Conflicting firing rules can be identified and dataflow between actors can be analysed to ensure progress and/or determinism.
If non-determinism is required by the programmer, the language provides a non-deterministic choice, whose resolution is left to the responsibility of the programmer. 
Finally, non-determinism as introduced by the use of a dynamic scheduler can also be
reduced by identifying deterministic schedules at compile time, using a model-checker as in ~\cite{Ersfolk2011}. 

Lohstroh et al.~\cite{LLDetAct19} recently proposed a deterministic actor language. The key ingredient for determinism is the \emph{logical timing} of messages based on a protocol which combines physical and logical timing to ensure determinism. 
The solution relies on the tagging of messages with a timestamp and determination of an arbitrary (deterministic) order for identical timestamps. To the best of our knowledge, there is no proof of correctness of the scheduling protocol.  


There is a significant adoption of dataflow programming models in the industry. One of the most prominent example is
TensorFlow~\cite{Abadi2016}, an efficient and flexible  framework for machine learning  based on  stateful dataflow graphs. 
Despite the dataflow inspiration, there is no determinism result for TensorFlow and more generally no high-level mechanism to restrict race-conditions;  programming artefacts equivalent to mutexes have to be used to restrict non-determinism~\cite{TensorFlow2017}.

      
\begin{strongpoints}{DPN}
\item[\PLUS] Generic programming model which fits many use-cases (ML, Signal Processing, HPC, \dots).
\item[\PLUS] Deterministic variants of DPN can be designed (e.g. \cite{LLDetAct19} that relies on timestamps and arbitrary deterministic ordering to provide deterministic reactors).
\item[\MEH] Tailored for task parallelism.
\item[\MINUS] Non-determinism needs to be skimmed away by programmer. 
\item[\MINUS] Static analysis needed to guarantee determinism, deadlock-freeness, etc.
\end{strongpoints}

  In the setting of CAL, each computing entity is sometimes called an actor.
  However, actors were already introduced back in the eighties~\cite{Agha86-book,AkkaBook} for a relatively similar notion. The main difference is that the actors in CAL are associated with a dataflow semantics, while more traditional actors, presented in the following section, are based on a sequential semantics.

\subsection{Actors and Active objects}
\label{sec:actors}
\edef\secnumact{\arabic{subsection}}

An actor \cite{Agha86-book,AkkaBook} is a mono-threaded entity that communicates with others by asynchronous 
message sending. In particular two actors cannot access the same memory location, which prevents data races.
Active objects~\cite{CSUR2017} unify the notions of actor and object (in the sense used in object-oriented programming), as they give to each actor an object 
type and replace message passing by \emph{asynchronous method invocation}: an active object
communicates by calling methods on other active objects, asynchronously.

In active object languages,
the result of an asynchronous method call is a future \cite{ABCL1994} (see \cref{def:fut}). 
An actor program is generally designed quite similarly to a sequential program except  message sending that triggers asynchronous computations. When a reply to a message is expected, either the programmer must implement an explicit callback or rely on a future if possible.
The internal behaviour of an actor is deterministic, and the only source of non-determinism  is either the concurrent sending of messages to the same destination, or a non-deterministic scheduling of message treatment (e.g. if the mailbox is not FIFO). The absence of multi-threading inside an actor and the fact that each data is handled by a single actor prevents  
 data races.

 ProActive is one of the first full-fledged active object languages.
 It uses active objects to implement a Java library made of distributed objects that communicate asynchronously. The ASP (Asynchronous Sequential Processes)
 calculus~\cite{CHS:POPL04,CH-book} formalises the 
 ProActive framework. It has been used to prove partial determinism properties that can be summarised as follows 1) \emph{If the dependence graph between active objects is a tree then the execution is deterministic}; 2) \emph{The result of a program execution is entirely determined by  the ordered list
of the identifiers of the caller in the request queue of each active object}.
 The work on ASP also relates deterministic active objects to Kahn process networks~\cite{CH-book}.
 The partial determinism of ASP comes from the actor model but also the FIFO service of requests in the model.
Interestingly, this result somehow show that futures do not compromise the determinism properties of programs provided they are accessed in a blocking manner; somehow justifying the fact that similar futures have been successfully introduced in Lustre without loosing determinism.
 %
 %
 The determinism properties of ProActive have been used to design a fault-tolerance protocol for active objects~\cite{BCD:PPoPP07}.
 More recently~\cite{DBLP:conf/ifm/HenrioJP20}, these determinism properties have been revisited in a more general setting, enabling a restricted form of cooperative scheduling, a type system ensuring determinism for active objects have also been designed.

AmbientTalk~\cite{DedeckerCMDM-ecoop06} was created based on the E Programming Language~\cite{Miller05concurrencyamong} which implements an actor model with a communicating event-loop.
It targets embedded systems and uses asynchronous future access (see below). Asynchronous reaction on future resolution and more generally on events is a prominent source of non-determinism in AmbientTalk.

 Creol \cite{johnsen03nik,johnsen07sosym} and the languages that inherit from it, JCobox~\cite{schafer2010jcobox}, ABS 
\cite{JHSSS10}, and Encore \cite{ref:encore15},  rely on \emph{cooperative scheduling} allowing the single execution thread of the active object to interrupt the service of one request and start (or recover) another one, but only at program points specified by the programmer. This approach clearly introduces more non-determinism than ASP for two reasons: 1) Cooperative scheduling introduces interruption points, generally depending on external events like the resolution of a future, but they are placed by the programmer who can control the sources of non-determinism. 2) There is no predefined order on the scheduling of pending tasks and the scheduler is thus by nature non-deterministic.
More recently, partial determinism properties of ABS have been exploited to reproduce a given execution~\cite{TJS:FASE20:ReproduceABS}; the approach is based on the recording of non-deterministic events that include request service and local scheduling. The fact that in active objects the set of non-deterministic decision in the scheduling of an active object is very small makes this approach feasible and scalable.
%

In a more industrial setting, Akka~\cite{haller2009scala,AkkaBook} is a scalable library for actors on top of Java and Scala.  The Akka documentation encourages programmers to use asynchronous reaction on futures. This introduces a significant source of non-determinism, suggesting that (partial) determinism is not one of the objectives of the language.
Software transactional memory has also been combined with
actors in the context of Akka to perform speculative computation and improve performance while ensuring the absence of data races in
such actors~\cite{HSHMF:AkkaSTM2013}.

\begin{strongpoints}{Actors and Active Objects}
\item[\PLUS] Mesh particularly well with object-oriented languages.
\item[\PLUS] Easy to use, and distributed as library in mainstream languages.
\item[\MEH] Tailored for task parallelism.
\item[\MEH] Can scale well with sufficient engineering efforts. 
\item[\MINUS] Race conditions exist, in particular due to communications.
\end{strongpoints}


 \subsection{Bulk Synchronous Parallel}
\label{subsec:bsp}
 
 Bulk Synchronous Parallel (BSP) \cite{Valiant1990}  targets data-parallelism where parallelism appears when the same task is performed on different data.
BSP  is a parallel execution model that defines algorithms as a sequence of 
supersteps, each made of three phases: computation, communication, and synchronisation.
BSP is  adapted to the programming of data-parallel applications and requires
the strong synchronisation of all computing entities.
A BSP program is typically made of a piece of code that is a step of computation and terminates by a synchronisation operation that coordinates all processes.
Interactions between BSP processes occur through communication primitives sending messages
or performing one-sided  Remote Direct Memory Access (RDMA) operations.

 While task parallelism, addressed for example by actors, is more general because it is always possible to split data and then launch different tasks, data-parallel programming models are more specialised and can run programs much more efficiently provided the application is inherently data-parallel. Typically, BSP allows the programmer to easily write a program that performs a heavy computation on a huge matrix and run it efficiently on high-performance computers, provided the computation can be split into independent treatments on parts of the matrix.

BSP features 
predictable 
performance and absence of deadlocks under simple hypotheses. The fact that computation is separated into independent entities makes it very easy to design deterministic algorithms in BSP.
Only concurrent writes on the same memory address, in the implementations of BSP that support it, can be a source of non-determinism.
Recently, \cite{HHLS-coordination2018} integrated BSP and actors,
leading to a model that features both task and data parallelism
with limited non-determinism.
Additionally,  it is possible to design a static analysis that checks deadlock-freedom for BSP programs~\cite{JAKOBSSON2017535,DBLP:conf/sac/Dabrowski18}.

Compared to the races that can appear in active objects, the way concurrent communication is handled in BSP is very different: in active objects there is no data race but communication races are inherent to the model and intertwined with computations. In BSP, the computation part is completely race-free and races that appear during communication steps can always be made deterministic, e.g. by prioritising a process. As the synchronisation step involves no computation and is bounded, a reordering of communications is always safe (but might induce  delays in  the synchronisation step).

There is a quite wide adoption of the BSP programming model in the industry.
For example, Pregel is a graph processing framework adapted to computation over large graphs inspired by BSP and developed by Google~\cite{Pregel2010}.


\begin{strongpoints}{BSP}
\item[\PLUS] Simple to learn for sequential programmers.
\item[\PLUS] Easy to ensure deadlock freedom and other properties, by analysis or by restricting the language. 
\item[\PLUS] Has a cost model.
\item[\MEH] Tailored for coarse-grained data parallelism.
\item[\MINUS] Limited programming model.
\end{strongpoints}

\subsection{Algorithmic Skeletons}
\label{sec:algskel}

Algorithmic Skeletons \cite{Gonzalez-Velez:2010:SAS:1890754.1890757} are a high-level parallel programming model introduced by Cole \cite{Cole:1991:ASS:128874}. 
Skeletons express parallel composition patterns through a composition language that assembles basic (sequential) blocks.
A pattern is a high-level construct that composes some operations and may introduce parallelism. The pattern can provide data-parallelism when using a farm pattern that  instantiates several instances of the same skeleton to work on several independent pieces of data, or task-parallelism when instantiating a pipeline where each stage runs in parallel (on a different input).
 Composition of skeletons is realised by using a set of classical parallel programming patterns proposed to the programmer, typically \emph{map}, \emph{reduce}, \emph{pipeline}, \emph{farm}, \emph{divide and conquer}, etc.

The expressive power of algorithmic skeletons comes from the variety of patterns proposed to the programmer,  while their efficiency is obtained by specific scheduling techniques used by the execution environment. 
 The Map-reduce pattern~\cite{Buono20102095} is one of the most massively used instances of algorithmic skeletons, it is dedicated to data-intensive computations and is successful because of the simplicity of its usage and the efficiency of its implementations. Dynamic and speculative versions of polyhedral-model based optimisation use dynamic information to chose between skeleton patterns for final code generation~\cite{sukumaranrajam:tel-01251748}.

While there is no focus on determinism properties, most instances of algorithmic skeleton programs are by nature deterministic because each sequential skeleton is independent and in general the composition operation preserves determinism.
Basically non-determinism could appear during some composition operations that  gather results coming from different computations. For example the reduce phase of a Map-reduce computation could be dependent on the order of production of the gathered results. But very often reduction phases are programmed so that they are not sensitive to the order of computation and skeleton programs are often deterministic in practice. 
Overall, ensuring determinism only requires to ensure that some operations are commutative.
 To the best of our knowledge no formal specification of determinism properties or no deterministic skeleton runtime has been proposed yet.

Calcium~\cite{CL07:Calcium} is an algorithmic skeleton library implemented over the active object library ProActive, enabling  a large-scale distributed  execution of a variety of skeletons. The separation into actors provided by the active object pattern naturally fits well with the notion of independent, hierarchically organised skeletons. No property of determinism has been identified in this context despite the use of the ProActive library that features partial determinism properties

\begin{strongpoints}{Algorithmic Skeletons}
\item[\PLUS] Excellent composition.
\item[\PLUS] High-level/good optimisation opportunities.
\item[\MINUS] Very limited programming model.
\end{strongpoints}

\subsection{Linear Types}
\label{sec:lineartypes}

Linear types are a language feature enhancing the capabilities of the underlying language through a richer type system. Linear types
provide efficient low level control over copy and aliases~\cite{walker} and data race freedom \emph{by construction}.
%
%
For this purpose, linear types enforce ``linear'' uses of resources, where
resources must always be used exactly once. These resources can therefore
not be shared or copied by other functions, restricting
concurrent accesses and data races, and preventing
various programming errors such as use-after-free.
For instance, if a database handle is used linearly, it cannot be shared
between different functions, and thus cannot be accessed concurrently.
Since it cannot be shared, when the database handle is closed it cannot
be used any more.

This initial idea has been expanded to numerous contexts. In
functional programming, it birthed two families of type systems,
linearity-by-functions
\cite{walker,DBLP:journals/pacmpl/BernardyBNJS18}
where functions must themselves use their argument linearly and
linearity-by-kinds
\cite{DBLP:conf/tldi/MazurakZZ10,DBLP:conf/popl/TovP11,DBLP:conf/icfp/Morris16}
where resources are given a usage mode
which might be linear.
Finally, unique types~\cite{DBLP:conf/plilp/BarendsenS95}
enforce that resources have a single reference to them which enables
aggressive optimisations for purely functional parallel languages~\cite{DBLP:conf/pldi/HenriksenSEHO17} while preserving
data race freedom.
Several systems provide formal proofs of non-concurrent accesses
for resources used linearly \cite{walker,DBLP:journals/pacmpl/RadanneST20}.

To handle imperative programming with mutable states,
linear types have been extended with complementary notions such as
regions~\cite{DBLP:conf/popl/LucassenG88,DBLP:conf/oopsla/BocchinoADAHKOSSV09}
and ownership~\cite{DBLP:conf/popl/BoylandR05}.
Regions allow to delimit an area of memory that functions are allowed to use.
Parallelism can then be limited to operations on distinct regions,
thus preventing aliasing.
Ownership relaxes this notion by allowing a function
to temporarily ``borrow'' a resource, which can be used in limited ways
(for instance, it cannot free it).
This concept has been widely used in Rust \cite{rust}, along
with other imperative and functional
languages~\cite{DBLP:conf/pldi/GrossmanMJHWC02,DBLP:journals/pacmpl/RadanneST20}.
\cite{DBLP:journals/corr/abs-1903-00982,DBLP:journals/pacmpl/0002JKD18}
provides formal proofs
that Rust ensures data race freedom for linear resources.

Finally, capabilities~\cite{DBLP:conf/ecoop/BoylandNR01} and
typestates~\cite{DBLP:conf/pldi/DeLineF01,DBLP:conf/oopsla/AldrichSSS09}
combine linearity with object-oriented programming to
 precisely control  what objects can or can't do,
such as ``this channel can send messages now''.
They have notably been used in parallel languages with actors
such as Encore~\cite{DBLP:conf/sfm/BrandauerCCFJPT15}
to prevent data races.

In all these cases, linear types serve as support for the underlying programming
model by encoding additional checks in the type system.
This allows to statically check numerous properties
for safety (absence of data races or race-conditions),
efficiency (additional no-aliasing guarantees),
or to adapt to different use-cases, e.g. for data-parallelism~\cite{rayon},
task-parallelism~\cite{actix}, or
lock-free data-structures~\citep{DBLP:conf/ecoop/CastegrenW17}.
The flip slide is that such checks need to be encoded in the discipline
of type systems, which requires great care by the programming language
designers, and occasionally exposes programmers to concepts difficult to
understand.

\begin{strongpoints}{Linear Types}
\item[\PLUS] Enable as much control as C with POSIX Threads, with additional safety.
\item[\PLUS] Feedback on potential races through typing errors. 
\item[\PLUS] Composes well with other paradigms and multipurposes languages.
\item[\MINUS] Steep learning curve.
\item[\MINUS] Safety properties requires careful design of the compilers.
\end{strongpoints}
  
\section{Discussion}
\label{sec:summary}

\subsection{Summary: How to Limit Non-determinism?}
\label{sec:classification}

This section provides a general discussion over the models we have presented earlier, and identifies how they help dealing
with races and limit non-determinism. 
We identify the following mechanisms used to obtain determinism, which can be formulated as questions of the form:
``Does (and if yes, how?) the approach ...: 
\begin{enumerate}
\item Separate computation from communication; 
\item Rely on sequential computation as much as possible;
\item Limit the potential interleaving of communications (or synchronisations); 
\item Protect or prevent access to shared data.''
\end{enumerate}

The goal of this section is neither to compare the various techniques nor
the programming models, but rather to present the landscape of approaches
in broad strokes and to hint at gaps in the existing ecosystems.
It also aims to provide a guide to adequately choose among programming
constructs and models with respect to the desired guarantees.

The rest of this section summarises the approaches we studied in this survey
and identifies the techniques they use to address the four questions.
\cref{tab:classif} provides a ``bird's eye view'' of this summary,
organised by properties and approaches.

\subsubsection{Separate Computation from Communication}
\label{summary:compcom}

The first mechanism we identify to limit non-determinism is the ability the programmer has to specify
the interplay between computations and communications.

A first possibility is for the programmer herself to
describe in a \emph{fine-grained} manner
how communications are interleaved with computations.
The most extreme examples are manual use of threads, as well as linear types.
KPN, DPN, and actor models all allow the programmer to mix computations and communications in a fine-grained fashion, albeit
differently from one another.
In KPN, reads are blocking and all communication channels are point-to-point, which is a combination of properties that guarantees determinism effectively.
In DPN and actor models, communication, performed through multiple writer mailboxes do not interrupt computations, which provides partial determinism. The races between different messages is a source of non-determinism but it also brings  flexibility and reactivity. 

Other approaches use \emph{coarse-grained} communications,
typically at the frontier of intersteps.
In these approaches, no communication can occur while the program is performing computations on data, and all data are supposedly available when the computation starts.
This is typically the case for BSP, where the communications are written down by the programmer.
This is also the case for synchronous dataflow languages, where the programmer only describes data dependencies which are then translated by the compiler into supersteps of the form ``get input data / compute / output data''.

Finally, some approaches derive communication \emph{automatically}, without
explicit knowledge provided by the programmer.
In the polyhedral model and in algorithmic skeletons,
no information about data exchanged by threads is ever given by the programmer. All communications and synchronisations are inferred from the program's data dependencies.





\subsubsection{Rely on Sequential Computation as Much as Possible}
\label{summary:seqcomp}

The second identified mechanism to limit non-determinism is to actually have the programmer write as much of her program's behaviour in a sequential manner.
Each programming model can be characterised by  a \emph{sequential quantum} which varies depending on the desired property of the programming model.

First, the quantum can span the full program, like in polyhedral model, letting the compiler parallelise it, with the help of annotations.

In many other approaches, the quantum is identified by programming constructs.
For instance,
BSP programs are segmented into ``supersteps'', each of which contains
sequential code.
In algorithmic skeletons, sequential code is contained in each computation
function (also called muscles).
In actor models, DPN or KPN, sequential code is contained
in entities whose granularity is left to the programmer.
Such programming constructs can also be implicit, such as regions for linear
types.
%

Finally, in synchronous dataflow languages, the actions described inside a block
are individually sequential. Any of them can be fired concurrently during
the execution of a block.
Non-determinism is handled by sequentializing the description at compile-time.


When ``sequential quantum'' are identified, non-determinism can be limited by controlling the way these quantum interact and can be interleaved. We first review how to control interleaving of messages below before focusing on sharing data in the following section.

\subsubsection{Limit Interleaving of Communications}
\label{summary:interleaving}

The more communications are constrained, the less variability is possible in terms of scheduling, hence the less non-determinism will appear. 
At one extreme, in algorithmic skeletons, as well as polyhedral or
synchronous dataflow programs,
the programmer has no control over communications and the scheduling
is entirely \emph{statically computed}.
The exact method of communication depends on the approach.
This can be slightly relaxed: for instance, explicit communications in Rust can still be safe thanks to the usage of linear types.

In BSP, all the outputs of a sequential quantum are communicated ``at once'' using a \emph{barrier} when their new values have all been computed.
All communications can be seen as deterministically performed at the inter-superstep level.

In DPN and KPN, communications are mixed with computations.
This is reinforced by the blocking-read semantics of \emph{channels}:
when no data is available on an input port, the actor is blocked, and no other form of communication can happen.
Unlike KPN, DPN's communication is not always deterministic.

Actors also feature blocking-reads, but on a mailbox that can be filled by several other actors. This can lead to non-determinism due to races when sending messages to the mailbox. There is no consensus on the use of a partial ordering on communications and some models rely on asynchronous communications while some others enforce a causal ordering of messages.

\subsubsection{Protect or Prevent Access to Shared Data}
\label{summary:shareddata}
Sharing data is an efficient way to exchange information between processes but it is a frequent source of non-determinism and bugs.
One solution to prevent races is the control of the access to this shared data.

The first solution is to only do atomic operations on the shared data.
For instance, Software Transactional Memory-based approaches limit the non-determinism introduced by data races by ensuring that transactions are executed atomically, without conflicting access to data.
This is very efficient thanks to the lock-free approach but the order of the transactions is still in general non-deterministic.
By extension, we can consider BSP as atomic, as communications are done in an atomic step with respect to computations thanks to barriers.

The second solution is to rely on single assignment with blocking reads
(\cref{def:sassign}), notably
Futures (\cref{def:fut}) and FIFO channels (\cref{def:FIFO}). Indeed the cells of the FIFO channels used in synchronous approaches, as well as DPN and KPN, can be seen as memory cells accessed through single assignment with blocking reads.
More generally, linear types also ensure determinism and safety for the access to shared data.
Futures and linear types do not prevent race conditions between threads but they
prevent all forms of data races.
%


\begin{table}[!t]
  \centering
  \newcommand*\rot[1]
  {\multicolumn{1}{l}{\rlap{\rotatebox{70}{\makecell[l]{#1}}}}}
  \newcommand*\notrot[2]
  {\multicolumn{1}{b{#1}}{\centering #2}}
  \scalebox{0.8}{
  \begin{tabular}
    {|r|m{30mm}|m{26mm}|m{38mm}|m{41mm}|}
    \multicolumn{1}{r}{}
    & \notrot{30mm}{
      Separation of computation and communication\\
      \cref{summary:compcom}
      }
    & \notrot{26mm}{
      Span of sequential computations\\
      \cref{summary:seqcomp}%
      }
    & \notrot{38mm}{
      Means to limit interleaving of communications\\
      \cref{summary:interleaving}
      }
    & \notrot{41mm}{
      Potential ways to protect or prevent access to
      shared data\\
      \cref{summary:shareddata}
      }
    \\\hline
    Seq C
    & NA
    & Full program
    & NA
    & NA
    \\\hline
    \makecell[r]{C+POSIX \\ Threads}
    & Manual
    & Manual
    & Mutex (Manual)
    & Atomic (Manual, Semaphore)\newline
      Future (Manual)
    \\\hline\hline
    \makecell[r]{KPN\\\cref{sec:kpn}}
    & Explicit and\newline fine-grained
    & ``Process''
    & Dynamic channels
    & FIFO channels
    \\\hline
    \makecell[r]{StaticDF\\\cref{sec:sdf}}
    & Explicit and\newline fine-grained
    & ``Process''
    & Compiled channels
    & FIFO channels
    \\\hline
    \makecell[r]{SyncDF\\\cref{subsec:sync}}
    & Automatic
    & Block
    & No control, Compiled communications
    & FIFO channels \newline
      Future (\cite{lustrefutures:emsoft12})
    \\\hline
    \makecell[r]{Polyhedral\\model\\\cref{sec:seq}}
    & Automatic
    & Full program
    & No control, Compiled communications
    & Single assignment with blocking reads \newline
      Linear types (\cite{DBLP:conf/pldi/HenriksenSEHO17})
    \\\hline\hline
    \makecell[r]{DPN\\\cref{sec:dpn}}
    & Fine-grained
    & Actor
    & Dynamic channels
    & FIFO channels
    \\\hline
    \makecell[r]{Actors\\\cref{sec:actors}}
    & Explicit and\newline fine-grained
    & Actor
    & Mailboxes
    & Future \newline
      Linear types (\cite{DBLP:conf/sfm/BrandauerCCFJPT15})
    \\\hline
    \makecell[r]{BSP\\\cref{subsec:bsp}}
    & Explicit and\newline coarse-grained
    & Superstep
    & Barriers
    & Atomic (Barriers)
    \\\hline
    \makecell[r]{Algorithmic\\skeletons\\\cref{sec:algskel}}
    & Automatic
    & Computation\newline function
    & No control, Compiled communications
    & Many approaches
    \\\hline
    \makecell[r]{Linear types\\\cref{sec:lineartypes}}
    & Fine-grained
    & Region
    & Control by typing
    & Single assignment (Types)\newline Futures (\cite{rust:tokio})
    \\\hline
  \end{tabular}}
\caption{Summary of our classification of parallel programming models.
  The four columns describe the various techniques used
    to enable determinism, described in \cref{sec:classification}.
}
  \label{tab:classif}
\end{table}

\medskip

All approaches described in the survey propose in essence a trade-off over these three major properties.  In order to
enhance their possibilities, future work may combine the mechanisms described in~\cref{tab:classif}. Programming models that are less constrained rely on a clever combination of the four mechanisms identified above: 
combining two mechanisms provides a more relaxed programming discipline than the strict application of one mechanism.
For instance, actor languages use futures to compensate for the strict separation of data. Each actor is sequential but global execution relies on asynchronous messages that provide task parallelism, and causal ordering can be used for message communications to make the model more deterministic.

Similarly, in the case of synchronous languages, several properties are applied with the additional constraint brought by the synchronous hypothesis that requires every computation step to fit inside a given timeslot. From this point, full determinism is guaranteed because data do not need to be protected as shared data is automatically copied. No synchronisation is necessary because the synchronous hypothesis ensures the availability of data. Finally, communications are implicit and occur between steps and are thus separated from computations.

While we attempted to map out the various mechanisms for determinism in this section, most of the possible combination remain unexplored.

\subsection{Strengths of the different Approaches presented in this Survey}

\label{sec:assessment}

In addition to~\cref{tab:classif}, we now identify characteristics
that we deem fundamental when coming to choosing a language or
modeling approach. These are based on our experience as programmers
and language designers. For each characteristic, we give some
approach examples that particularly seem to address them: 
\begin{itemize}[leftmargin=*]
\item {\bf Efficiency} is the ability to execute programs
  with the best performances, according to a set of benchmarks. We believe that efficiency is only meaningful if crucial properties, like determinism or absence of data races, are still guaranteed.
  To achieve efficiency, different approaches are appropriate for different applications. {\bf The polyhedral model} (\cref{sec:seq}) achieves great
  efficiency for fine grain parallelism with regular dependencies
  though automatic parallelisation. {\bf Linear types}
  (\cref{sec:lineartypes}) provide ``zero-cost'' abstraction, which
  allows programmers to write the most efficient code while ensuring determinism without additional runtime check.
  Finally {\bf Bulk
    Synchronous Parallel} (\cref{subsec:bsp}) come with a cost model allowing programmers to evaluate the efficiency of the parallelisation pattern used.
  
\item {\bf Safety} is the ability of the programming model to
  enable additional safety properties (worst case execution time, non
  aliasing, absence of use after free, \ldots). {\bf Synchronous
    dataflow} (\cref{subsec:sync}) enables precise dependence
  computation, which provides worst-case estimations and static sizes
  for buffers. {\bf Linear types} (\cref{sec:lineartypes}) guarantee
  determinism as a side-effect of the strict typing discipline
  which also guarantees non-aliasing and absence of use after-free, among others
  safety properties.
  
\item {\bf Ease of use} is the ``entry cost'' for programmers
  used to mainstream languages.
  In this category, {\bf Actors} (\cref{sec:actors}) are of notable interest
  as they provide concurrent semantics over a simple extension of
  Object-Oriented programming, making it familiar to all programmers
  used to this paradigm. {\bf Algorithmic skeletons}
  (\cref{sec:algskel}) propose high-level operators which are easy to
  compose and usable within any language.
\end{itemize}





\section{Conclusion}

In this survey we proposed a comprehensive review of data race free and deterministic existing parallel programming models. We proposed a classification based on different choices used by each programming model to guarantee determinism. All these criteria are summarised in \cref{tab:classif}. We also discuss additional efficiency, safety and usability characteristics of these approaches, based on our own experience. 
More important than the classifying table, \cref{sec:classification} analyses and classifies the approaches used by the different languages to limit non-determinism, providing insight on the strength of each language and the way the existing approaches can be combined. 
We believe our classification can be used  to chose among programming models, both as a programmer choosing an existing programming language, or as a language designer creating a safe and efficient parallel programming language. The bird-eye view provided by the summarising table can also be used to provide ideas for new combinations of models and/or approaches.

This survey also allows us to study precisely different programming methodologies that are often considered by different communities and thus not directly compared in details. Indeed the approaches presented here belong to object-oriented programming, synchronous languages, compilation techniques, functional programming, or high-performance computing, which are rarely if ever discussed together. 

Finally, we would like to highlight a possible research direction that can be identified as a missing combination according to our review:
the combination of futures or linear types with techniques providing structural constraints on programs, such as dataflow approaches.
Indeed, futures and linear types would provide flexibility in these strongly structured languages, without threatening the powerful safety properties provided
by the underlying dataflow structure.


{\small
\bibliographystyle{alpha}

\bibliography{biblio,biblioludo}
}

\end{document}